\documentclass[10pt]{article}
\usepackage{graphicx} 
\usepackage{amsmath}
\usepackage{hhline}
\usepackage{stackengine}
\usepackage{amsthm}
\usepackage{blkarray}
\usepackage{mathrsfs}
\usepackage[margin=1.7in]{geometry}
\usepackage{amssymb}
\usepackage{mathtools}
\usepackage{xcolor}
\usepackage{algorithm}
\usepackage{algpseudocode}
\usepackage{bbold}
\usepackage{cite}
\usepackage{latexsym}
\usepackage{booktabs}
\usepackage{bm}
\usepackage{textcomp}
\usepackage{subfigure}
\usepackage[toc,page]{appendix}
\usepackage{url}
\usepackage{hyperref}
\usepackage{cleveref}
\usepackage{multicol}
\usepackage{multirow}
\usepackage{tikz}
\usetikzlibrary{3d}
\usepackage{tikz-3dplot}
\usepackage{pgfplots}

\let\oldtextit\textit 
\renewcommand\emph[1]{\oldtextit{\color{RoyalBlue}#1}}
\definecolor{RoyalBlue}{cmyk}{1, 0.50, 0, 0}
\theoremstyle{definition}
\newtheorem{example}{Example}[section]
\newtheorem{definition}[example]{Definition}
\newtheorem{thm}[example]{Theorem}
\newtheorem{remark}[example]{Remark}

\newtheorem{lemma}[example]{Lemma}

\newcommand{\RR}{{\mathbb R}}

\newcommand{\cA}{{\mathcal A}}

\newcommand{\mainfilecheck}[1]{0}

\newcommand{\true}{\textnormal{\textsc{True}}}
\newcommand{\false}{\textnormal{\textsc{False}}}

\makeatletter
\def\@settitle{\begin{center}%
  \baselineskip13\p@\relax
    \Large
\@title
  \end{center}%
}
\makeatother
\title{Certified algebraic curve projections by path tracking}

\author{Michael Burr
  \and
  Michael Byrd
  \and
  Kisun Lee
}

\newcommand{\Addresses}{{
  \bigskip
  \footnotesize

  Michael Burr, Michael Byrd, and Kisun Lee, \textsc{School of Mathematical and Statistical Science, Clemson University, 220 Parkway Drive, Clemson, SC 29634}\par\nopagebreak
  \textit{E-mail addresses}, \url{burr2@clemson.edu}, \url{mbyrd6@clemson.edu}, \url{kisunl@clemson.edu}

}}

\date{}

\begin{document}

\maketitle

\begin{abstract}

We present a certified algorithm that takes a smooth algebraic curve in $\mathbb{R}^n$ and computes an isotopic approximation for a generic projection of the curve into $\mathbb{R}^2$. Our algorithm is designed for curves given implicitly by the zeros of $n-1$ polynomials, but it can be partially extended to parametrically defined curves.  The main challenge in correctly computing the projection is to guarantee the topological correctness of crossings in the projection. Our approach combines certified path tracking and interval arithmetic in a two-step procedure: first, we construct an approximation to the curve in $\mathbb{R}^n$, and, second, we refine the approximation until the topological correctness of the projection can be guaranteed. We provide a proof-of-concept implementation illustrating the algorithm.

\end{abstract}

\section{Introduction}
In numerical algebraic geometry, curve tracking algorithms were originally introduced for solving polynomial systems using homotopy continuation, see, e.g., \cite{Sommese:2005,Bertini:2013} and the references therein.  More recently, the study of the connectivity structure defined by homotopy paths has opened new avenues of research, e.g., \cite{Sommese:2001,GaloisDuff:2022}.  With the recent development of efficient certified homotopy continuation methods in \cite{xu2018approach,guillemot2024validated,duff2024certified}, the computations of these connectivity structures become proofs.  The current paper continues this approach, using homotopy continuation and interval arithmetic to produce certified approximations to curves in arbitrary dimensions as well as their generic projections to $\mathbb{R}^2$.

Suppose $C$ is a regular algebraic curve without self-intersections in $\mathbb{R}^n$ for some $n\geq 3$. Let $x\in\mathbb{R}^n$ approximate a point $x^\star$ on the curve.  We present two certified algorithms for computations with this curve.  First, we provide an algorithm to compute an isotopic approximation to the component of the curve $C$ which contains $x^\star$ along with a tubular neighborhood containing this component of the curve $C$.  Second, if $\pi:\mathbb{R}^n\rightarrow\mathbb{R}^2$ is a generic projection, then we compute an isotopic approximation to the image of the component of the curve $C$ which contains $x^\star$ under the projection $\pi$.  In our calculations, the curve $C$ is usually given implicitly as the zero set of a system of $n-1$ polynomials, but our approach also applies to open curves given parametrically by $x=\gamma(t)$, where $\gamma$ consists of $n$ univariate polynomials.

For our first algorithm, we combine and develop ideas from the certified homotopy continuation algorithms \cite{guillemot2024validated,duff2024certified} with the certified curve tracking algorithm \cite{martin2013certified}.  This results in a certified approximation to the curve which can be locally refined as needed.  We note that one of the significant differences between our approach and that of \cite{martin2013certified} is that we do not use a global lower bound on the allowed step-size, so our algorithm is guaranteed to terminate with the correct output on all regular curves.

In our second algorithm, we introduce ideas from \cite{Byrd:2023} to correctly approximate intersections in the image of the curve after a generic projection to $\mathbb{R}^2$.  We observe that, due to a dimension argument, a generic projection $\pi:\mathbb{R}^n\rightarrow\mathbb{R}^m$ with $m>2$ introduces no intersections.  On the other hand, projections to $\mathbb{R}^2$ typically introduce intersections. Processing these intersections is the main challenge in the construction of an approximation.  In particular, the issue is that the projection of an approximation may not be an approximation of the projection, as errors in the approximation may introduce or miss self-intersections.

\subsection*{Outline}

In Section \ref{sec:background}, we review the necessary background for the paper from the certified computation literature.  In Section \ref{sec:path_tracking_algorithm}, we present the path tracking algorithm under study in this paper, see Algorithm \ref{algo:certified_curve_tracking}.  We apply this algorithm in Section \ref{sec:approximation} to develop Algorithm \ref{algo:certified_plane_curve}, which computes certified approximations to generic projections of curves.  Finally, Section \ref{sec:implementation} introduces a prototype Julia implementation with illustrative examples.

\section{Background}\label{sec:background}
Suppose that $C$ is a regular algebraic curve without self-intersections in $\mathbb{R}^n$ given by the common zeros of a polynomial system $F$ consisting of $n-1$ polynomials in $n$ variables.  Here, regular means that the Jacobian of $F$ is always full rank.  We also observe that the self-intersection restriction does not exclude closed curves.  

The algorithms in this paper compute certified piecewise-linear approximations to the curve $C$ and its generic projections.  In this context, a certified algorithm is one that provides a proof of correctness for its output, and, for our problems, the output is correct if (1) the approximation and curve are close in Hausdorff distance and (2) there is an ambient isotopy deforming the curve to the approximation.  In all of our algorithms, we consider polynomial systems and inputs with coefficients in $\mathbb{Q}$ because they can be represented exactly on a computer, avoiding having to address how arbitrary real numbers are represented.

There is an extensive history of curve tracking algorithms, but there are relatively fewer certified path tracking algorithms, see, e.g., \cite{kearfott1994interval,Kim:2004,PV:2004,PV:2007,BCGY:2012,martin2013certified,Hoeven2015,katsamaki2023ptopo,Byrd:2023}.  Several certified approaches \cite{kearfott1994interval,Kim:2004,martin2013certified,Hoeven2015} are based on certifying predictor-corrector techniques.  In addition, \cite{PV:2004,PV:2007,BCGY:2012,Byrd:2023} use subdivision-based techniques until the curve (or curves) have simple local behavior.  The paper \cite{katsamaki2023ptopo} solves the related problem of tracking parametrically defined curves without converting the system to an implicit system.  For the special problem of path tracking in homotopy continuation, there are several approaches that certify the constructed curve, see, e.g., \cite{Smale:1993,Beltran:2008,Burgisser:2011,beltran2012certified,beltran2013robust,hauenstein2014posteriori,Hauenstein:2016,xu2018approach,guillemot2024validated,duff2024certified}.  Of these algorithms, it seems that \cite{guillemot2024validated,duff2024certified} may be among the most practical.  In the current paper, we combine ideas from \cite{guillemot2024validated,duff2024certified} to create a certified path tracking algorithm whose correctness is based on interval arithmetic and the Krawczyk method.  An important feature of our approach, which is lacking from many of the previous algorithms, is that the approximation can be locally refined after the completion of the computation.

\subsection{Interval arithmetic}
Interval arithmetic is a method of computing with intervals instead of single numbers.  One advantage of using intervals to represent numbers is that conservative calculations in interval arithmetic can be used to quantify numerical errors.  Another use of interval arithmetic is that larger intervals can be used to study the behavior of functions over regions.  By performing arithmetic using intervals, our algorithms ensure reliable results.  

For a given arithmetic operator $\odot$, such as addition or multiplication, and intervals $I$ and $J$, we define 
\begin{equation}\label{eq:interval_arithmetic}
    I\odot J=\{x\odot y\mid x\in I, y\in J\}
\end{equation}
For instance, $[a,b]+[c,d]=[a+c,b+d]$; see, e.g., \cite{moore2009introduction} for more details.  We denote the collection of real intervals by $\mathbb{IR}$, and those with rational endpoints by $\mathbb{IQ}$. For a function $f:\mathbb{R}^n\rightarrow \mathbb{R}$ and an $n$-dimensional interval vector $I=(I_1,\dots, I_n)\in \mathbb{IQ}^n$, we define an \emph{interval extension} $\square f(I)$ of $f$ over $I$ to be an element of $\mathbb{IR}$ satisfying 
\[\square f(I)\supset \{f(x)\mid x\in I\}.\]
An interval extension $\square f(I)$ for a specific $f$ and $I$ is not unique.  There may be several formulas for interval extensions of $f$ over $I$ which result in different over-approximations, e.g., for a polynomial function $f\in \mathbb{Q}[X_1,\dots, X_n]$, an interval extension may be obtained by replacing variables by intervals and applying \Cref{eq:interval_arithmetic}. We define the interval extension $\square F(I)$ for a system of equations $F:\mathbb{R}^n\rightarrow \mathbb{R}^m$ by extending each coordinate function individually.

For an interval $I\in \mathbb{IR}$, we denote the maximum magnitude of an interval by  $\|I\|:=\max\limits_{x\in I}|x|$. For interval vectors and matrices, we introduce formulas mimicking the induced norms. More precisely, for an interval vector $I=(I_1,\dots, I_n)$, we define $\|I\|=\max\limits_{i=1,\dots, n}\|I_i\|$. In addition, for an interval matrix $M\in\mathbb{IR}^{n\times m}$, we define $\|M\|=\max\limits_{A\in M}\max\limits_{x\in \mathbb{R}^n}\frac{\|Ax\|}{\|x\|}$ where $\|x\|=\max\limits_{i=1,\dots,n}|x_i|$.


\subsection{The Krawczyk test}
The Krawczyk method is a central tool in certified path tracking as success with it proves the existence and uniqueness of a solution within a region.  The test combines a generalized Newton's method-type computation with interval arithmetic to prove that a specified region contains a unique solution.  Krawczyk \cite{krawczyk1969newton} introduced an interval operator that refines a compact convex region in order to isolate roots of nonlinear systems.  Moore \cite{moore1977test} demonstrated how to use this method to prove the existence of a solution to a nonlinear system. Building on this, Rump \cite{rump1983solving} extended the approach to establish the uniqueness of the solution.  Using this test, we define when an approximate solution is certified by the Krawczyk method.

\begin{definition}[\cite{guillemot2024validated}]\label{def:approximate_solution}
    Given a polynomial system $F:\mathbb{R}^n\rightarrow\mathbb{R}^n$ and some $\rho\in (0,1)$, a point $x\in \mathbb{R}^n$ is called a \emph{$\rho$-approximate solution} to $F$ if there is an $n\times n$ invertible matrix $A$ and a constant $r>0$ such that  \begin{equation}\label{eq:krawczyk_test}
-AF(x)+\left(I_n-AJF(x+rB)\right)rB\subset r\rho B, 
    \end{equation}  
    where $B=[-1,1]^n$ is the unit interval box in $\RR^n$ and $JF$ denotes the Jacobian of $F$.  For a $\rho$-approximate solution, \cite[Theorem 2.1]{guillemot2024validated} implies that a unique solution $x^\star$ to $F$ exists in $x+rB$, with $\|x-x^\star\|\leq \rho r$, and we say that $x^\star$ is the \emph{associated solution} to $x$.
\end{definition} 

When $x$ is a $\rho$-approximate solution, the quasi-Newton map $g(x)=x-AF(x)$ is a $\rho$-Lipschitz continuous function.  The constant $\rho$ represents how accurate $x$ must be.  In other words, as $\rho$ gets smaller, the distance between $x$ and $x^\star$ must be smaller for the containment in \Cref{eq:krawczyk_test} to be satisfied. The following test uses interval arithmetic to confirm the containment in \Cref{eq:krawczyk_test}.

\algrenewcommand\algorithmicrequire{\textbf{Input}:}
\algrenewcommand\algorithmicensure{\textbf{Output}:}

\begin{algorithm}[ht]
	\caption{KrawczykTest}
 \label{algo:Krawczyk-test}
\begin{algorithmic}[1]
\Require  
\begin{itemize}
    \item A polynomial system $F=\{f_1,\dots, f_n\}\subset\mathbb{Q}[X_1,\dots, X_n]$,
    \item a point $x=(x_1,\dots, x_n)\in \mathbb{Q}^n$ (or $\mathbb{IQ}^n$),
    \item a positive real number $r\in \mathbb{Q}$,
    \item an $n\times n$ invertible matrix $A\in\mathbb{Q}^{n\times n}$ (or $\mathbb{IQ}^{n\times n}$),
    \item $\rho\in (0,1)$.
\end{itemize}
\Ensure A boolean.
\State {Set $K=-\frac{1}{r}A F(x)+\left(I_n-A\square JF(x+rB)\right)B$.}
\State {Return $\|K\|< \rho$.}
 \end{algorithmic}
 \end{algorithm}

\begin{remark}
Although we focus on real curves in this paper, it is natural to extend interval arithmetic to $\mathbb{C}^n$ by considering $\mathbb{C}^n$ as $\mathbb{R}^{2n}$ and modifying the operations appropriately.  In addition, the Krawczyk test and definition of an approximate solution can be adapted to the complex case, see, e.g., \cite{Analytic2019,Breiding:2023,guillemot2024validated,duff2024certified} for more details.
\end{remark}

\subsection{Certified homotopy path tracking}

\emph{Homotopy path tracking} is a method to find a solution to a system of nonlinear equations.  In order to solve the system $F:\mathbb{C}^n\rightarrow\mathbb{C}^n$, the method uses another system $G:\mathbb{C}^n\rightarrow\mathbb{C}^n$ whose solutions are known in advance and a homotopy $H(x;t):\mathbb{C}^n\times [0,1]\rightarrow\mathbb{C}^n$ such that $H(x;0)=G(x)$ and $H(x;1)=F(x)$.  If $F$ and $G$ both have finitely many nonsingular solutions, then solutions to $F$ can be found by starting at solutions to $G$, and tracking solution paths from $t=0$ to $t=1$.  \emph{Certified} homotopy path tracking algorithms guarantee their correctness by constructing a compact region that contains the homotopy path. This compact region has the added property that, for any fixed $t^\ast\in [0,1]$, there is only one solution to $H(x;t^\ast)$ in the region.  There are two main methods for constructing this compact region.  The approaches in \cite{beltran2012certified,beltran2013robust,hauenstein2014posteriori} use Smale's alpha theory to certify their solution paths.  On the other hand, \cite{kearfott1994interval,martin2013certified,Hoeven2015,xu2018approach,duff2024certified,guillemot2024validated} use interval arithmetic-based approaches for certification.  In the current paper, we adapt the methods of \cite{duff2024certified,guillemot2024validated} from homotopy path tracking to more general path tracking.

\section{Curve tracking algorithm}\label{sec:path_tracking_algorithm}

We introduce a curve tracking algorithm for a regular curve in $\mathbb{R}^n$. Our approach is inspired by the homotopy tracking algorithms from \cite{duff2024certified,guillemot2024validated}.  The main distinction between homotopy continuation and path tracking is that, in homotopy continuation, there is a distinguished variable $t$, while in path tracking, the best that one can do is to have a locally distinguished variable.  We illustrate how interval arithmetic and the Krawczyk method are exploited to track the curve in a certified manner. 

Let $C$ be a regular curve in $\mathbb{R}^n$ given by the real loci of an algebraic variety $\mathbf{V}(c_1,\dots, c_{n-1})$ defined by $n-1$ polynomials $c_i\in \mathbb{Q}[X_1,\dots, X_n]$ for $i=1,\dots, n-1$.  We note that we use the notation $C$ for both the algebraic variety $\mathbf{V}(c_1,\dots,c_{n-1})$ as well as the polynomial system $\{c_1,\dots, c_{n-1}\}\subset\mathbb{Q}[X_1,\dots, X_n]$. 
Then, the Jacobian $JC$ of the curve is an $(n-1)\times n$ matrix. We denote the intersection of the curve $C$ and the plane $X_n=a$ by $C_a$.  Moreover, we note that when $a\in\mathbb{Q}$, $C_a:\mathbb{R}^{n-1}\rightarrow \mathbb{R}^{n-1}$ is a square polynomial system in $\mathbb{Q}[X_1,\dots, X_{n-1}]$.  On the other hand, for a point $x=(x_1,\dots, x_n)\in\mathbb{R}^n$, we let $x_{-n}=(x_1,\dots, x_{n-1})$ denote the point in $\mathbb{R}^{n-1}$ consisting of the first $n-1$ coordinates of $x$.


\subsection{Unitary transformation}

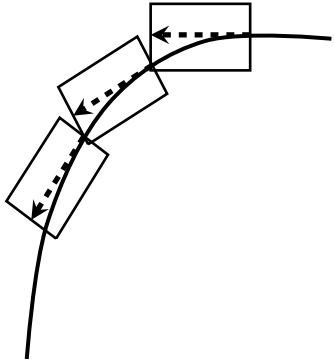
\begin{figure}
    \centering
    \begin{tikzpicture}[scale=6]
    
        
        \draw[color=black, line width = 1.5] plot[smooth, tension = .6] coordinates{(.15, 0.1) (.195, 0.4) (0.33, 0.665) (0.545, 0.805)  (.825, .81)  }; 

        \draw[black, line width = 1] (.215,.367) -- (.33,.553) -- (.223,.635) -- (.105,.45) -- (.215,.367);
        \draw[black, line width = 1] (.286,.576) -- (.461,.688) -- (.395,.815) -- (.22,.703) -- (.286,.576);
        \draw[black, line width = 1]         (.4245,.74) -- (.645,.74) -- (.645,.8875) -- (.4245,.8875) -- cycle;
        
        \draw[line width=2pt,black,-stealth, dashed] (0.2765, 0.594)--(0.16, 0.4085) node[anchor=south west]{};	
        \draw[line width=2pt,black,-stealth, dashed] (0.428, 0.7515)--(0.253, 0.6395) node[anchor=south west]{};	
        \draw[line width=2pt,black,-stealth, dashed] (0.645, 0.81875)--(0.4245, 0.81875) node[anchor=south west]{};	
        
\end{tikzpicture}
    \caption{A schematic illustration of curve tracking with rotation.  At each step, the curve is rotated so that the tangent vector is $e_n$.}
    \label{fig:curve_rotation}
\end{figure}

The first step to adapt homotopy path tracking to general curve tracking is to update the distinguished direction.  Unlike a solution path in a homotopy, which maintains a consistent direction as $t$ increases, a general curve may change its direction over the path. A straightforward remedy is to rotate the curve to temporarily align its tangent vector with a specified direction at each step of the algorithm, see \Cref{fig:curve_rotation}. We introduce a unitary transformation designed for this purpose in \Cref{algo:transform}.

\algrenewcommand\algorithmicrequire{\textbf{Input}:}
\algrenewcommand\algorithmicensure{\textbf{Output}:}

\begin{algorithm}[ht]
	\caption{UnitaryTransformation}
 \label{algo:transform}
\begin{algorithmic}[1]
\Require  
\begin{itemize}
    \item A regular curve $C=\{c_1,\dots, c_{n-1}\}\subset\mathbb{Q}[X_1,\dots, X_n]$, and
    \item a point $x=(x_1,\dots,x_n)\in \mathbb{Q}^n$ approximating a point on $C$.
\end{itemize}
\Ensure \begin{itemize}

    \item A transformed curve $\hat{C}$,
    \item a transformed point $\hat{x}$, and
    \item unitary matrices $U$ and $V^\star$.
\end{itemize} 
\State {Compute the SVD of $JC(x)=U\Sigma V^\star$\\
Set $\hat{C}=U^\star C(V(X))$.\\
Set $\hat{x}=V^\star x$.\\
Return $(\hat{C},\hat{x}, U, V^\star)$.}

 \end{algorithmic}
 \end{algorithm}

For a regular curve $C$ and point $x$, let $U\Sigma V^\star$ be the SVD of $JC(x)$ and $\hat{x}\vcentcolon=V^\star x$ be the corresponding image of $x$ under the rotation $V$.  We construct the rotated curve $\hat{C}(\hat{x})\vcentcolon=U^\star C(V\hat{x})$, by rotating the curve $C$. 
We observe that
    $$
    J\hat{C}(\hat{x})= U^\star JC(x)V=U^\star U\Sigma V^\star V=\Sigma.
    $$
We note that $\Sigma$ is an $(n-1)\times n$ matrix whose last column consists of all zeros. Hence, $\ker J\hat{C}(\hat{x})=\langle e_n\rangle$. When $x$ and $y$ are points on the curve $C$, $V^\star x$ and $V^\star y$ are also points on the curve $\hat{C}$. 

\begin{remark}
We note that \Cref{algo:transform} requires the computation of the singular value decomposition. For rigorous computation, we perform this computation using interval arithmetic so $U$, $\Sigma$, and $V^\star$ are all represented by interval matrices with arbitrarily small intervals.  In this case, the diagonal structure and the column of zeros of $\Sigma$ are both maintained.  Since these matrices can be made with arbitrarily small intervals, any subsequent calculation can be made as precise as needed. 
\end{remark}


To avoid excessive use of $\hat{C}$ and $\hat{x}$, we always assume that $C$ and $x$ have been replaced by $\hat{C}$ and $\hat{x}$, respectively. Then, we combine this approach with \Cref{def:approximate_solution} to arrive at the following: for a regular curve $C\subset \mathbb{R}^n$, we say that a point $x=(x_1,\dots, x_n)\in \mathbb{R}^n$ is a \emph{$\rho$-approximate solution} to $C$ if $x_{-n}$ is a $\rho$-approximate solution to $C_{x_n}$. In this case, an associated solution $x^\star$ is a point on the curve $C$ where the last coordinate satisfies $X_n=x_n$. We begin by adapting the prediction-correction schemes from \cite{duff2024certified,guillemot2024validated} to our setting.  Then, we discuss the subtleties in connecting individual approximations.

\subsection{Curve prediction}

The core idea of our certified curve tracking algorithm is to construct a series of regions $\{I_i\}_{i=1,\dots,m}$ that begin from one point on (or near) the curve and cover a portion of the curve.
The Krawczyk test is applied to these regions to ensure that only one arc of $C$ is contained within each $I_i$ and the curve behaves nicely within each $I_i$. To construct such a tube, we use a numerical approximation of the curve (referred to as a \emph{prediction}) and create a region around this approximation. If the Krawczyk test from \Cref{algo:Krawczyk-test} passes for this region, we proceed to the next step of tracking.

Two methods for curve prediction have been recently proposed for certified homotopy continuation \cite{duff2024certified,guillemot2024validated}.  In \cite[Algorithm 2]{duff2024certified}, the authors exploit the predictor-corrector method, see, e.g., \cite[Section 2.3]{sommese2005numerical}, for a prediction of the curve.  On the other hand, \cite[Section 6.3]{guillemot2024validated} employs a higher-order approximation of the curve obtained from a point on the curve and a tangent vector. The Taylor model, see \cite[Section 9.3]{moore2009introduction}, is applied to construct a curved interval box containing this higher-order approximation.  One of the notable differences between these two prediction methods is the use of the step size and a corrector. The predictor algorithm in \cite{duff2024certified} requires the desired step size as input. 
In contrast, \cite{guillemot2024validated} derives an appropriate step size directly from the data.

For a curve $C$ and a point $x$, we denote a prediction of the curve by $X(\eta):[0,\infty)\rightarrow \mathbb{R}^{n-1}$ such that $(X(0),x_n)=x$. The algorithm to produce such a prediction may be chosen by the user. For a radius $r$, when the algorithm from \cite{duff2024certified} is chosen, we denote this algorithm by $\text{Predictor}(C,x,r,h)$ with a desired step size $h$, or $\text{Predictor}(C,x,r)$ when the prediction algorithm from \cite{guillemot2024validated} is chosen.  Since the choice of prediction is a user-defined choice, in our images we draw enclosing boxes for the prediction tube to illustrate the structure even though the true prediction tube may be curved.

\subsection{Refinement}

Since it is difficult to access an exact point on the curve, we must apply the prediction algorithm to a point close to the curve.  The more accurate a given point is, the more precise the prediction is and the more efficient the corresponding tracking is.  The refinement step replaces a given approximation by a more accurate approximation, leading to better behavior for the predictor.  \Cref{algo:meta_refine} performs this refinement. In particular, it takes a $\rho$-approximate solution $x$ with an associated solution $x^\star$ to $C$ as an input and returns a point $x$ and $r$ such that all points in $x+(\boldsymbol{0}_{n-1}\times r[-1,1])$ are $\tau$-approximate solutions to $C$ for an input $\tau\in (0,1)$. In this case, a point $y\in x+(\boldsymbol{0}_{n-1}\times r[-1,1])$ and its associated solution $y^\star$ share the same last coordinate, that is, $y_n=y_n^\star$.  In \Cref{sec:curve_tracking_algorithm}, we illustrate why refining the region $x+(\boldsymbol{0}_{n-1}\times r[-1,1])$ and replacing $x$ with a better approximation are useful in our algorithms.

\begin{algorithm}[ht]
	\caption{RefineSolution~\cite[Algorithm 2]{guillemot2024validated}}
 \label{algo:meta_refine}
\begin{algorithmic}[1]
\Require  
\begin{itemize}
    \item A regular curve $C=\{c_1,\dots, c_{n-1}\}\subset\mathbb{Q}[X_1,\dots,X_n]$,
    \item a $\rho$-approximate solution $x=(x_1,\dots, x_n)$ to $C_{x_n}$ with corresponding radius $0<r<1$ and an $(n-1)\times (n-1)$ invertible matrix $A$, and
    \item a constant $\tau\in (0,1)$.
\end{itemize}
\Ensure A $\tau$-approximate solution $x$ with a corresponding radius $\tilde{r}$ and an $(n-1)\times (n-1)$ invertible matrix $\tilde{A}$.
\State{Set $x_{-n}=(x_1,\dots, x_{n-1})$}
\State{Set $I=[-1,1]\in\mathbb{IQ}$}
\State{Set $\tilde{r}=r$}
\State{Set $\tilde{A}=A$} 
\While{$\text{KrawczykTest}(C_{x_n+\tilde{r}I}, x_{-n}, \tilde{r}, \tilde{A}, \tau)=\false$}
\If{$\|A\square C_{x_n}(x_{-n})\|\leq \frac{1}{8}(1-\rho)\tau \tilde{r}$}
\State{Set $\tilde{r}=\frac{1}{2}\tilde{r}$}
\Else
\State{Set $x_{-n}=x_{-n}-AC_{x_n}(x_{-n})$}
\State{Set $x=(x_{-n},x_n)$}
\EndIf
\State{Set $\tilde{A}=JC_{x_n}(x_{-n})^{-1}$}
\EndWhile
\While{$2\tilde{r}\leq 1$ and 

{$\!\!\!\text{KrawczykTest}(C_{x_n+\tilde{r}I}, x_{-n}, 2\tilde{r}, \tilde{A}, \tau)=\true$}}\label{line:while_loop_in_Refine}
\State{Set $\tilde{r}=2\tilde{r}$.}
\EndWhile
\State{Return $x,\tilde{r},\tilde{A}$.}
 \end{algorithmic}
 \end{algorithm}

As \Cref{algo:meta_refine} refines a $\rho$-approximate solution to an approximate solution with a smaller constant, we use this algorithm to refine the computed regions for a curve. Refining these regions is needed when we need a finer approximation with a smaller tubular neighborhood. This feature plays an important role in computing the isotopic approximation of a projection of a curve in \Cref{sec:approximation}.

\Cref{algo:meta_refine} is heavily inspired by \cite[Algorithm 2]{guillemot2024validated}. For completeness, we provide a correctness and termination statement for the algorithm. A detailed analysis can be found in \cite[Section 4.2]{guillemot2024validated}. 
\begin{lemma}
\Cref{algo:meta_refine} returns a $\tau$-approximate solution and terminates within finitely many iterations.
\end{lemma}
\begin{proof}
    Suppose the algorithm does not terminate.  The input $x$ is a $\rho$-approximate solution, so the map $g(x)=x-AC_{x_n}(x_{-n})$ is a $\rho$-Lipschitz continuous function by \cite[Theorem 2.1]{guillemot2024validated}.  Therefore, $x_{-n}$ converges to the fixed point, i.e., an exact point on the curve.  Therefore, with sufficiently high precision, $\|A\square C_{x_n}(x_{-n})\|$ becomes arbitrarily close to zero.  Hence $\tilde{r}$ must halve infinitely many times.  Since $\tilde{r}$ becomes arbitrarily small, we may assume it produces negligible errors in the computations.

    As in Algorithm \ref{algo:Krawczyk-test}, let
    $$K=-\frac{1}{\tilde{r}}A C_{x_n}(x_{-n})+\left(I_n-A\square JC_{x_n}(x_{-n}+\tilde{r}B)\right)B.$$
    Whenever $\tilde{r}$ is halved, the first term is bounded from above by $\frac{1}{8}(1-\rho)\tau<\tau$.  On the other hand, $\|I_n-A\square JC_{x_n}(x_{-n}+\tilde{r}B)\|\rightarrow 0$ as $\tilde{r}$ approaches zero as $\square JC_{x_n}(x_{-n}+\tilde{r}B)\rightarrow JC_{x_n}(x_{-n}^\star)$ and $\tilde{A}\rightarrow JC_{x_n}(x_{-n}^\star)^{-1}$.  Therefore, the Krawczyck test eventually succeeds.
    
    Therefore, the first loop must terminate within finitely many iterations. The second loop enlarges the radius in each iteration. It terminates after fewer than $-\log_2(\tilde{r})$ iterations.
    %
   %
\end{proof}

The first while loop reduces the radius at each iteration, potentially resulting in a narrow region. Tracking using a narrow interval box can be ineffective because a narrow box restricts the shape of the curve.  The second while loop enlarges the radius to avoid this.

The difference between Lines 6 and 9 is that, in Line 6, $\square C_{x_n}(x_{-n})$ is computed via an over-approximation to guarantee the inequality.  On the other hand, in Line 9, any point within $\square C_{x_n}(x_{-n})$ can be used for $x_{-n}$, if the precision is sufficiently high.

The main difference between Algorithm \ref{algo:meta_refine} and \cite[Algorithm 2]{guillemot2024validated} appears in Lines 5 and 14.  In these lines, the Krawczyk test is performed on an entire interval at once by using $x_n+\tilde{r}I$.  The idea is to introduce an interval for the slice of $C$ being considered.  By using such an interval, the Krawczyk test verifies the existence of a solution for every instance in the interval at once.  This refinement is needed in our main curve tracking algorithm, Algorithm \ref{algo:certified_curve_tracking}.

\begin{remark}
In \cite[Algorithm 2]{guillemot2024validated}, the precision of the computations is adjusted during the computation.  We observe that this implies that for high enough precision, the computation succeeds.  We leave the details of adaptive precision to the reader.
\end{remark}

\subsection{Stopping criterion}

We consider two stopping criteria for curve tracking. For a regular curve $C$ in $\mathbb{R}^n$ and a $\rho$-approximate solution $x$ to $C$ with the radius $r$, we suppose that the user provides a compact region of interest $D$ that contains $x$ such that the curve does not meet $\partial D$ transversely. Tracking terminates whenever we prove that the curve is closed or the curve crosses the boundary of $D$.  We denote these criteria by $\text{StoppingCriterion}(D,x,C)$, and provide details below.

If a newly constructed interval region $I$ intersects the boundary so that the end of the curve is guaranteed to be outside of $D$, then we conclude that the curve has reached $D$.  If it is not possible to guarantee that the end of the curve is outside of $D$, further refinement of the approximation is needed to decide if the curve crosses $\partial D$.  Since the curve intersects the boundary transversely, this decision is possible.  If the refinement of $I$ does not cross the boundary, then we continue tracking the curve.

To conclude that the curve $C$ is closed, we check if $I$ intersects an interval vector $J$ (which do not form a pair of sequential steps). If they intersect, we refine both independently. If the refinement of the exit point of the curve in $J$ is contained in $I$, then we conclude that the curve is closed. If the refinement of $J$ does not intersect $I$ (or its refinement), we continue tracking the curve. 

\subsection{Curve tracking algorithm}\label{sec:curve_tracking_algorithm}

We propose \Cref{algo:certified_curve_tracking} to track a regular curve $C$ in $\mathbb{R}^n$ starting with the point $x$.  We assume that $x$ has been refined enough so that $\hat{x}$ is a $\rho$-approximate solution after the initial unitary transformation.  We recall that the singular value decomposition uses interval arithmetic, so there may be roundoff error in its computation.  Hence, we require that the precision has been increased to make $\|VV^\star- I_n\|$ as small as needed.  Similarly, the computation of any inverses in the algorithm can be made precise enough so that any subsequent error is as small as necessary.  

\begin{algorithm}[ht]
	\caption{CertifiedCurveTracking}
 \label{algo:certified_curve_tracking}
\begin{algorithmic}[1]
\Require  
\begin{itemize}
    \item A regular curve $C=\{c_1,\dots, c_{n-1}\}\subset\mathbb{Q}[X_1,\dots, X_n]$,
    \item an approximate solution $x\in\mathbb{Q}^n$ to the curve $C$,
    \item a constant $r>0$,
    \item a constant $h>0$,
    \item a compact region $D$ in $\mathbb{R}^n$,
    \item constants $\rho\in(0,\frac{1}{2}]$ and $\tau\in(\frac{1}2,1)$
\end{itemize}
\Ensure {\begin{itemize}
    \item A set of interval regions $\mathcal{I}$, and
    \item a $\rho$-approximate solution $x$ of $C$ outside $D$ or a certificate that $C$ is a closed curve.
\end{itemize}}
\State{Set $(\hat{C},\hat{x}, U, V^\star)=\text{UnitaryTransformation}(C,x)$, where $\hat{x}=V^\star x$}.
\State{Set $A= J\hat{C}_{\hat{x}_n}(\hat{x}_{-n})^{-1}$.}
\State{Set $\mathcal{I}=\emptyset$.}
\While{$\text{StoppingCriterion}(D,x,C)=\false$}
\State{Set $h=\frac{5}{4}h$.}
\State{Set $m=0$.}
\Repeat
\State{Set $(\hat{x}, r, A) =\text{RefineSolution}(\hat{C},\hat{x},r,A,\frac{\rho}{2^m})$.}
\State{Set $m=m+1$.}
\State{Set $x=V\hat{x}$.}\label{line:rotate_back}
\State{Set $(\tilde{C},\tilde{x}, \tilde{U}, \tilde{V^\star})=\text{UnitaryTransformation}(C,x)$.}\label{line:unitary_transformation}
\State{Set $\tilde{A}= J\tilde{C}_{\tilde{x}_n}(\tilde{x}_{-n})^{-1}$.}
\Until{$\text{KrawczykTest}(\tilde{C},\tilde{x},r,\tilde{A},\rho)=\true$}
\State{Set ($\hat{x},r,A)=\text{RefineSolution}(\tilde{C},\tilde{x},r,\tilde{A},\rho)$}\label{line:6}
\State{Set $U=\tilde{U}$}
\State{Set $V=\tilde{V}$}
\State{Set $\hat{C}=\tilde{C}$}
\State{Set $\hat{X}(\eta)=\left\{\begin{array}{ll}
    \hat{x}_{-n} & \text{ if }\eta<0 \\\text{\quad (backward extension)}  \\
    X(\eta)=\text{Predictor}(\hat{C},\hat{x},r,h) & \text{ if } \eta\geq 0\\ \text{\quad(predicts the forward curve)}
\end{array}\right.$.}\label{line:retract}
\While{$\text{KrawczykTest}(\hat{C}_{\hat{x}_n+[-r\rho,h]},\hat{X}([-r\rho,h]),r,A,\tau)=\false$}\label{line:10}
\State{$h=\frac{h}{2}$.}
\State{Set $X(\eta)=\text{Predictor}(\hat{C},\hat{x},r,h)$.}
\EndWhile
\State{Set $\mathcal{I}=\mathcal{I}\cup\{V\cdot(\hat{X}([-r\rho,h]),\hat{x}_n+[-r\rho,h])^\top\}$.}
\If{$I\cap J\ne\emptyset$ for any non-adjacent $I,J\in \mathcal{I}$ (except if the StoppingCriterion holds).}
\State{Break. Try again with a smaller $\rho$.}
\EndIf
\State{Set $\hat{x}=(X(h),\hat{x}_n+h)$.}\label{line:update}
\EndWhile
\State{Return $\mathcal{I}$ and $x$ or a conclusion that $C$ is a closed curve.}
 \end{algorithmic}
 \end{algorithm}

\begin{figure}
    \centering
    \begin{tikzpicture}[scale=6]
    \draw[color=black, line width = 1.5] plot[smooth, tension = .6] coordinates{
         (.195, 0.31) (0.33, 0.665) (0.545, 0.79) (.825, .857)  
    };

    \draw[black, line width = 1, rounded corners=0.02cm] 
        (.68,.718) -- (.68,.942);
    \draw[black, line width = 1, rounded corners=0.02cm] 
        (.38,.657) -- (.38,.882);
 
    \draw[color=black, line width = 1] plot[smooth, tension = .6] coordinates{
        (.38, .66) (.60, .71) (.68, .72)  
    }; 
    
    \draw[color=black, line width = 1] plot[smooth, tension = .6] coordinates{
        (.38, .88)  (.60, .93) (.68, .94)  
    };

         \draw[dashed, line width = 1] 
        (.38,.688) -- (.419,.725); 

    \draw[line width=1pt,black,-stealth, dashed] plot[smooth, tension = .6] coordinates{
        (0.38, .688) (0.29, .605)  (0.2, 0.47) 
    }; 

         \draw[color=blue,, line width = .5] 
        (.428,.638) -- (.443,.622); 
         \draw[color=blue,, line width = .5] 
        (.466,.675) -- (.481,.659); 
         \draw[color=blue,, line width = .5] 
        (.435,.63) -- (.474,.667); 
    \node[blue, inner sep=1pt] at (0.49, .62) {{\small$r\rho$}};

    \draw[ line width = 1] 
        (.464,.675) -- (.37,.783); 
    \draw[ line width = 1] 
        (.24,.415) -- (.15,.525); 
    \draw[ line width = 1] 
        (.464,.678) -- (.423,.64); 
    \draw[ line width = 1] 
        (.375,.783) -- (.333,.745); 

    \draw[color=black, line width = 1] plot[smooth, tension = .9] coordinates{
        (0.42, .6375) (0.34, .555) (0.238, .414)  
    }; 
    
    \draw[color=black, line width = 1] plot[smooth, tension = .9] coordinates{
        (0.334, .745) (0.25, .665) (0.15, .524)  
    }; 
       
    \draw[red, line width = 1.5,dotted] 
        (.424,.637) -- (.33,.745); 
    \draw[red, line width = 1.5,dotted] 
        (.24,.415) -- (.15,.525); 

    \draw[color=red, line width = 1.5,dotted] plot[smooth, tension = .9] coordinates{
        (0.42, .6375) (0.34, .555) (0.238, .414)  
    }; 
    
    \draw[color=red, line width = 1.5,dotted] plot[smooth, tension = .9] coordinates{
        (0.334, .745) (0.25, .665) (0.15, .524)  
    }; 
    \draw[color=red, line width = 2] plot[smooth, tension = .9] coordinates{
        (0.383, .723) (0.37, .711) (0.36, .701)  
    }; 
    
    \node[circle, fill=blue, inner sep=1pt] at (0.38, .688) {};

\end{tikzpicture}
    \caption{The blue dot represents refined $\hat{x}$ from \Cref{line:6} in \Cref{algo:certified_curve_tracking}. The dashed arrow describes the prediction $\hat{X}$. The red dashed box does not cover the curve completely. The solid red portion of the curve shows the small uncovered gap when the red dashed box is used. It is covered by the interval extended backward by $r\rho$. The solid black boxes represent actual intervals computed to cover the curve.}
    \label{fig:retract_box}
\end{figure}
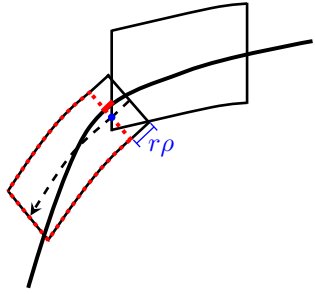

The main while loop in Algorithm \ref{algo:certified_curve_tracking} can be broken into two steps, the construction and certification.  In the first part, Lines 5-\ref{line:retract}, the parameters of the approximation are chosen via the unitary transformation.  In addition, the prediction is constructed in \Cref{line:retract}. 
We briefly focus on the loop in Lines 7-12.  In this loop, we generate a $\rho$-approximate solution to a rotated curve $\hat{C}$. Since we cannot guarantee that $\tilde{x}$ in \Cref{line:unitary_transformation} remains an approximate solution after the unitary transformation, we introduce several safeguards.  By using better approximations, the value in Line 1 of the Krawczyck test for $\tilde{C}$ approaches the corresponding value for $\hat{C}$.  Moreover, \Cref{algo:meta_refine} operates on a region of the form $x + r [-1,1]^n$, which ensures the disk of radius $r$ around $\tilde{x}$ remains within this region.  The vector $\hat{X}(\eta)$ in \Cref{line:retract} is constructed to extend backwards to cover the curve without any gaps. By evaluating $(\hat{X}(\eta),\hat{x}_n+\eta)$ over the interval $[-r\rho, h]$, the next step of the approximation completely covers the portion of the boundary of the previous interval region that contains the exiting curve.  Without extending the interval backwards, the intervals may not completely cover the curve (see \Cref{fig:retract_box}). Since refinement makes the initial tangent vector closer to the curve's tangent vector, by the interval-based Krawczyk test in Algorithm \ref{algo:meta_refine}, the curve does not leave the extended part.

The remaining lines certify the computation. \Cref{line:10} performs the Krawczyk test on an entire interval similar to the construction in Algorithm \ref{algo:meta_refine}.  If this fails, the step size is reduced and a new prediction is made.  The remaining lines add the new interval region to $\mathcal{I}$ and check if the non-neighboring boxes are intersecting.  If non-neighboring boxes intersect, then we cannot discern the behavior of the curve and must try again.  

When the algorithm succeeds, the tests and prior discussion imply that the result is correct.  What remains is to prove termination.  We now show that, with a small enough $\rho$, the algorithm succeeds.

\begin{lemma}
     \Cref{line:10} of \Cref{algo:certified_curve_tracking} returns $\true$ for some $h>0$.
\end{lemma}
\begin{proof}
From the Krawczyk test in \Cref{line:10}, let
$$    K(\eta)=-\frac{1}{r}A \hat{C}_{\hat{x}_n+\eta}(\hat{X}(\eta))\\+\left(I_n-A\square J\hat{C}_{\hat{x}_n+\eta}(\hat{X}(\eta)+rB)\right)B.$$ 
    We show that 
    $\|K([-r\rho,h])\|< \tau$
    for some $h>0$ after sufficiently many iterations of the while loop in  \Cref{line:10}.
    When $h$ converges to $0$, the vector $(\hat{X}([-r\rho,h]),\hat{x}_n+[-r\rho,h])$ converges to $(\hat{x}_{-n},\hat{x}_n+[-r\rho,0])$. Hence, there is some sufficiently small $h>0$ such that $\|K([-r\rho,h])\|< \rho<\tau$ because $\|K([-r,r])\|<\rho$ from Algorithm \ref{algo:meta_refine} in \Cref{line:6}.  Since this is a continuous function of $h$, the Krawczyk test succeeds when $h$ is sufficiently small.
\end{proof}

\begin{lemma}
    In \Cref{algo:certified_curve_tracking}, $\hat{x}=(X(h),\hat{x}_n+h)$ from \Cref{line:update} is a $\tau$-approximate solution to $\hat{C}$.
\end{lemma}
\begin{proof}
For a point $\hat{x}$ obtained from \Cref{line:6}, define
$$K(\eta)=-\frac{1}{r}A \hat{C}_{\hat{x}_n+\eta}(X(\eta))\\+\left(I_n-A\square J\hat{C}_{\hat{x}_n+\eta}(X(\eta)+rB)\right)B.$$ 
    From the second while loop, $\|K(h)\|< \tau$
    for any $\eta\in [-r\rho,h]$. This implies that $(X(\eta),\hat{x}_n+\eta)$ in \Cref{line:update} is a $\tau$-approximate solution.
\end{proof}

\begin{thm}
\Cref{algo:certified_curve_tracking} terminates.
\end{thm}
\begin{proof}
    Suppose that the sequence of points $x$ obtained in \Cref{line:rotate_back} converges to a point instead of satisfying the stopping criterion.  This means that the step size $h$ must be halved infinitely many times as we repeat the second while loop in \Cref{line:10}.  Let $$K(\eta,\lambda)=-\frac{1}{\lambda}A\hat{C}_{\hat{x}_n+\eta}(\hat{X}(\eta))+\left(I_n-A\square J\hat{C}_{\hat{x}_n+\eta}(\hat{X}(\eta)+\lambda B)\right)B.$$
    Then, we have $\|K([-r\rho,h],r)\|>\tau$ for infinitely many $h$.
    On the other hand, $\|K([-r\rho,r\rho],r)\|\leq \|K([-r,r],r)\|< \rho$ because of the results of Algorithm \ref{algo:meta_refine} from \Cref{line:6} in the algorithm. 
    
    We now show that in this case, $r$ also approaches zero.  For contradiction, suppose $r$ were bounded away from zero by $r_0$ from below at each iteration, i.e., $r\geq r_0$, then 
    $\lim\limits_{h\rightarrow 0}\|K([-r_0\rho,h],r_0)\|<\rho$.  This contradicts the fact that $$\|K([-r\rho,h],r)\|>\tau>\rho> \|K([-r\rho,r\rho],r)\|$$ for infinitely many $h$.
    Therefore, we may assume that $r$ becomes arbitrarily small due to the refinement in  \Cref{line:6}. 
    Then, whenever $r<1$ is returned from \Cref{line:6}, we have
    $\|K([-r,r],2r)\|>\rho$ from the Krawczyk test in the loop on \Cref{line:while_loop_in_Refine} of \Cref{algo:meta_refine}.
    Furthermore,
    since $\square J\hat{C}_{\hat{x}_n+[-r,r]}(\hat{X}([-r,r])+2rB)$ converges to $J\hat{C}_{\hat{x}_n}(\hat{x}_{-n})$ as $r$ converges to $0$, for any $0<\epsilon<1$, there is $r_0>0$ such that 
    \[\left\|I_n-A\square J\hat{C}_{\hat{x}_n+[-r,r]}(\hat{X}([-r,r])+2rB)\right\|<\epsilon\]
    whenever $r<r_0$.
    Thus, $K(\eta)$ is dominated by its first term, and whenever $r<r_0$ we have
    $$\left\|-\frac{1}{2r}A\hat{C}_{\hat{x}_n+[-r,r]}(\hat{X}([-r,r]))\right\|+\epsilon \geq \|K([-r,r],2r)\|>\rho.$$ This contradicts the inequality $\|K([-r,r],r)\|<\rho$.
\end{proof}

\begin{remark}
Algorithm \ref{algo:certified_curve_tracking} uses a global constant $\rho$ to guarantee the correctness.  It is possible to replace this global constant with a locally defined constant.  We leave these details to the interested reader.  In addition, for a detailed analysis of the homotopy path tracking algorithm with adaptive precision, see \cite[Section 5.3]{guillemot2024validated}.
\end{remark}

We say that the collection of all interval regions computed from \Cref{algo:certified_curve_tracking} is a \emph{tubular neighborhood} of the curve $C$ and denote this by $\mathcal{A}(C)$ (see \Cref{fig:tubular_nbd}).  We observe that by reducing $\rho$ as needed, we can make the Hausdorff distance between the approximation and the curve as small as desired.  In addition, if we wish to create an approximation to the curve itself, we may construct a path within the interval regions by connecting them sequentially.

\section{Approximating a plane curve}\label{sec:approximation}

We introduce a method for approximating a plane curve from the result of a generic projection $\pi : \RR^n \to \RR^2$ when $C$ is a regular curve.  In this case, generic means that the projection has no triple intersections, all of the crossings in the image are transverse, and there are no tangent vectors in the kernel of the projection.  Straight-forward dimension checks show that generic projections satisfy these three conditions.  To construct such an approximation, we start with a tubular neighborhood of the curve $C$ computed from Algorithm \ref{algo:certified_curve_tracking} and further refine until we can guarantee the correctness of the projection.  The main challenge with this approach is correctly handling self-intersections of $\pi(C).$

To achieve the correct topology, we use two tests.  One excludes intersections from regions and another detects crossings.  Suppose that $I_1$ and $I_2$ are two interval regions from $\mathcal{A}(C)$ such that $\pi(I_1)$ and $\pi(I_2)$ intersect.  This indicates that it is possible for the image of the curve to have an intersection within $\pi(I_1)\cap\pi(I_2)$.

We first discuss our exclusion test.  Let $(I_1,I_2)$ be the sequence of consecutive interval regions constructed in $\mathcal{A}(C)$ that goes between $I_1$ and $I_2$.  If $C$ is a closed curve, then both of the possible sequences can be used for $(I_1,I_2)$.  We exclude intersections between the curve in $I_1$ and $I_2$ if the direction of the curve doesn't change too much in $(I_1,I_2)$.  In particular, for each interval region $I$ in $(I_1,I_2)$, we evaluate the Jacobian $JC(I)$ and compute its numerical kernel $K$ using interval arithmetic.  In this case, $\pi(K)$ contains all tangent vectors to the curve $\pi(C)$ coming from $C\cap I$.  By the genericity assumption, for a sufficiently close approximation, $\pi(K)$ does not contain $0$.  We take the union of all $\pi(K)$ for $I$ in $(I_1,I_2)$.  If this union lies in an open half-space, then the image of the curve from $(I_1,I_2)$ cannot self-intersect in the projection as the projected curve is always traveling in the direction of the normal, and monotomic curves cannot self-intersect.  We call this test the \emph{half-space condition}.

If this test fails, then it is possible that there is a self-intersection of the curve in $\pi(I_1)\cap\pi(I_2)$.  In order to detect this crossing, we carefully approximate the image of the curve near $\pi(I_1)\cap\pi(I_2)$.  By studying the global behavior of the approximation, we can conclude if there is a self-intersection of the projection.  The main idea is to construct, for each of $I_1$ and $I_2$, a long rectangular region that contains the projection of the curve from either $I_1$ or $I_2$ such that the curve passes through the short ends of the rectangles.  If the rectangles intersect on their long sides only, then by the intermediate value theorem, the projections of the curves must cross.  Our test is detailed in Algorithm \ref{algo:strip_test}, see also Figure \ref{fig:crossing_rectangles}.

\begin{figure}[h]
    \centering
    \begin{tikzpicture}

		\draw[color=red, line width = 1.5] plot[smooth, tension = .8] coordinates{
			(0, 0) (1, 1.2) (2, 1.8) (3, 3.3) (4, 4)
		};
		
		\draw[color=blue, line width = 1.5] plot[smooth, tension = .8] coordinates{
			(4, 0) (3, 1.3) (2, 2.2) (1, 2.7) (0, 4)
		};

		\draw[color = red, line width = 1] (-0.1, 0) -- (-0.1, 0.5) -- (0.5, 0.5) -- (0.5, 0) -- (-0.1, 0);
		\draw[color = red, line width = 1] (0.2, 0.4) -- (0.2, 0.9) -- (0.9, 0.9) -- (0.9, 0.4) -- (0.2, 0.4);
		\draw[color = red, line width = 1] (0.5, 0.8) -- (0.5, 1.3) -- (1.2, 1.3) -- (1.2, 0.8) -- (0.5, 0.8);
		\draw[color = red, line width = 1] (1.1, 1.2) -- (1.1, 1.6) -- (1.6, 1.6) -- (1.6, 1.2) -- (1.1, 1.2);
		\draw[color = red, line width = 1] (1.5, 1.4) -- (1.5, 1.8) -- (1.8, 1.8) -- (1.8, 1.4) -- (1.5, 1.4);
		\draw[color = red, line width = 1] (1.7, 1.5) -- (1.7, 1.9) -- (1.9,1.9) -- (1.9, 1.5) -- (1.7, 1.5);
		\draw[color = red, line width = 1] (1.85, 1.6) -- (1.85, 2.1) -- (2.2,2.1) -- (2.2, 1.6) -- (1.85, 1.6);
		\draw[color = red, line width = 1] (2.0, 2.0) -- (2.0, 2.5) -- (2.6, 2.5) -- (2.6, 2.0) -- (2.0, 2.0);
		\draw[color = red, line width = 1] (2.3, 2.4) -- (2.3, 2.9) -- (2.9, 2.9) -- (2.9, 2.4) -- (2.3, 2.4);
		\draw[color = red, line width = 1] (2.5, 2.8) -- (2.5, 3.3) -- (3.1, 3.3) -- (3.1, 2.8) -- (2.5, 2.8);
		\draw[color = red, line width = 1] (2.8, 3.2) -- (2.8, 3.7) -- (3.7, 3.7) -- (3.7, 3.2) -- (2.8, 3.2);
		\draw[color = red, line width = 1] (3.2, 3.6) -- (3.2, 4.) -- (4., 4.) -- (4., 3.6) -- (3.2, 3.6);
		
		\draw[color = red, line width = .8, dashed] (-0.2, 0.5) -- (0.95, -0.46) -- (4.15, 3.53) -- (2.98, 4.47) -- (-0.2, 0.5);

		\draw[color = blue, line width = 1] (3.6, 0) -- (3.6, 0.4) -- (4., 0.4) -- (4., 0) -- (3.6, 0);
		\draw[color = blue, line width = 1] (3.3, 0.3) -- (3.3, 0.8) -- (3.9, 0.8) -- (3.9, 0.3) -- (3.3, 0.3);
		\draw[color = blue, line width = 1] (2.9, 0.6) -- (2.9, 1.1) -- (3.7, 1.1) -- (3.7, 0.6) -- (2.9, 0.6);
		\draw[color = blue, line width = 1] (2.7, 1) -- (2.7, 1.4) -- (3.5, 1.4) -- (3.5, 1) -- (2.7, 1);
		\draw[color = blue, line width = 1] (2.5, 1.3) -- (2.5, 1.7) -- (3.1, 1.7) -- (3.1, 1.3) -- (2.5, 1.3);
		\draw[color = blue, line width = 1] (2.3, 1.5) -- (2.3, 1.9) -- (2.9, 1.9) -- (2.9, 1.5) -- (2.3, 1.5);
		\draw[color = blue, line width = 1] (2.0, 1.75) -- (2.0, 2.1) -- (2.65, 2.1) -- (2.65, 1.75) -- (2.0, 1.75);
		\draw[color = blue, line width = 1] (1.8, 1.95) -- (1.8, 2.25) -- (2.45, 2.25) -- (2.45, 1.95) -- (1.8, 1.95);
		\draw[color = blue, line width = 1] (1.4, 2.05) -- (1.4, 2.5) -- (2.1, 2.5) -- (2.1, 2.05) -- (1.4, 2.05);
		\draw[color = blue, line width = 1] (1.1, 2.25) -- (1.1, 2.7) -- (1.75, 2.7) -- (1.75, 2.25) -- (1.1, 2.25);
		\draw[color = blue, line width = 1] (.55, 2.6) -- (.55, 3.) -- (1.3, 3.) -- (1.3, 2.6) -- (.55, 2.6);
		\draw[color = blue, line width = 1] (.3, 2.8) -- (.3, 3.4) -- (1.1, 3.4) -- (1.1, 2.8) -- (.3, 2.8);
		\draw[color = blue, line width = 1] (.1, 3.1) -- (.1, 3.7) -- (0.7, 3.7) -- (0.7, 3.1) -- (.1, 3.1);
		\draw[color = blue, line width = 1] (-0.1, 3.5) -- (-0.1, 4) -- (0.5, 4) -- (0.5, 3.5) -- (-0.1, 3.5);
		
		\draw[color = blue, line width = .8, dashed] (0.351, 4.2) -- (-0.6, 2.86) -- (3.7, -0.15)-- (4.64, 1.19)  -- (0.351, 4.2);
		
		\node[label=below:] at (1.75, .-0.1) {$R_1$};
		\node[label=below:] at (1.1, 4.0) {$R_2$};
	\end{tikzpicture}
    \caption{A schematic illustrating how self-intersections of the curves are detected.  The two rectangles contain the curve along their lengths and intersect on their lateral sides.  This guarantees a crossing of the curves.}
    \label{fig:crossing_rectangles}
\end{figure}

We note that Algorithm~\ref{algo:certified_curve_tracking} can be easily adapted to locally refine the tubular neighborhood $\cA(C)$.  Suppose that we wish to refine a consecutive sequence of tubes.  At each end of this sequence, there are points $x_1$ and $x_2$ that approximate the curve.  Without loss of generality, we assume that the direction in which we track $C$ passes $x_1$ first and then $x_2$.  By applying Algorithm \ref{algo:certified_curve_tracking} starting at $x_1$ and with small $\rho$ results in a refinement of the tubular neighborhood when the StoppingCriterion succeeds for the point $x_2$.

\begin{algorithm}[h]
	\caption{IntersectionCheck}
 \label{algo:strip_test}
\begin{algorithmic}[1]
\Require  
\begin{itemize}
    \item A regular curve $C=\{c_1,\dots,c_{n-1}\}\subset\mathbb{Q}[X_1,\dots,X_n]$
    \item A tubular neighborhood $\cA(C)$ obtained by Algorithm \ref{algo:certified_curve_tracking}
    \item A generic projection $\pi: \RR^n \to \RR^2$
    \item Interval regions $\{I_1, I_2\}$ with intersecting projections
\end{itemize}
\Ensure 
\begin{itemize}
    \item A boolean value indicating confirmation of a self-intersection.
\end{itemize}
\State{Set $p$ to be the center of $\pi(I_1) \cap \pi(I_2)$}
\State{Construct approximation $v_i$ to the image of the tangent vector at $p$ for $\pi(I_i)$.}
\If{$v_1$ and $v_2$ are parallel}
\State{Refine $I_1$ and $I_2$ and return \false}
\EndIf
\State{Set $r$ to be the maximum diameter of $\pi(I_1)$ and $\pi(I_2)$.}
\State{Construct rectangular region $R_i$ to be a rectangle centered at $p$ whose width is of length $2r$ in the direction $v_i^\perp$ and whose length is long enough so that the edges of length $2r$ are outside the other rectangle.}
\State{Refine $\cA(C)$ until every interval region intersecting $R_i$ has diameter at most $r$.}
\State{Starting at $I_i$, find the first interval region in the forward and backward directions of $\cA(C)$ that are completely outside $R_i$.  Set $(F_i,L_i)$ to be these consecutive sequence of regions.}
\If{Either $(F_i,L_i)$ intersects the lateral sides of $R_i$}
\State{Refine all interval regions in $R_i$ and return \false.}
\EndIf
\If{Each $(F_i,L_i)$ does not satisfy the half-space condition}
\State{Refine all interval regions in $R_i$ and return \false.}
\EndIf
\For{{\bf each} pair of interval regions $J_i\in R_i$}
\State{Set $K_i$ to be an interval box containing the numerical kernel of $JC(J_i)$}
\If{$\pi(K_1)$ and $\pi(K_2)$ contain a parallel vector}
\State{Refine all interval regions in $R_i$ and return \false.}
\EndIf
\EndFor
\State{Return \true}
\end{algorithmic}
\end{algorithm}



We observe that if Algorithm~\ref{algo:strip_test} returns \true, then the curve must have a self-intersection, as described above.  In addition, if there were multiple intersections in $R_1\cap R_2$, then at the extreme points, i.e., when the curves are farthest apart, their tangent vectors would point in the same (or opposite) directions.  This is prevented by Line 21 of Algorithm~\ref{algo:strip_test}.  This implies that success of Algorithm~\ref{algo:strip_test} implies that there is exactly one intersection in $R_1\cap R_2$.  We now prove that after sufficient refinements both intersections and non-intersections can be detected, see Figure \ref{fig:two_projections}.

\begin{lemma}
    Let $C$ in $\RR^n$ be a regular curve and $\pi: \RR^n \to \RR^2$ a generic projection. Let $\cA(C)$ be the tubular neighborhood of $C$ constructed with Algorithm \ref{algo:certified_curve_tracking}. For every self-intersection of $\pi(C),$ there exists a pair of interval regions $I_1$ and $I_2$ such that Algorithm~\ref{algo:strip_test} returns $\true$ after sufficient refinement.
\end{lemma}
\begin{proof}
Let $I_1$ and $I_2$ be two interval regions such that there is a point $p_i\in I_i$ such that both $p_i$'s project to the same point.  When $I_1$ and $I_2$ are sufficiently small, $v_1$ and $v_2$ are approximately the tangent vectors of the projected curve at the intersection point.  Since all self-intersections are transversal, the tangent vectors at the intersection are not parallel, so, after sufficient refinement, $v_1$ and $v_2$ cannot be parallel either.

After sufficient refinement, the vectors $v_1$ and $v_2$ approach the tangent vectors of the projected self-intersection, the angles of the sides of $R_1$ and $R_2$ become stable, so the aspect ratio of each $R_i$ becomes nearly fixed.  Observe that any curve can be described locally by $p_0+tv_0+O(t^2)$.  For each branch at the self-intersection point $p\rightarrow p_0$ and, for some $i$, $v_i\rightarrow v_0$.  Finally, since the size of $R_i$ also gets smaller with refinement, the error in the curve is eventually small enough so that it passes through the ends of $R_i$ of length $2r$.  Thus, after sufficient refinement, $(F_i,L_i)$ cannot intersect the lateral sides of $R_i$.

By the same argument as above, after sufficient refinement, the tangent vector to the projected curve is nearly constant.  In addition, the calculated numerical kernels are also almost constant.  Therefore, after sufficient refinement, the tangent vectors are all pointing in almost the same direction, so the half-space condition must hold.  Since the tangent vectors at the projected self-intersection are not parallel, this also implies that after sufficient refinement, the projected numerical kernels cannot contain parallel vectors.  

Thus, after sufficient refinement, Algorithm~\ref{algo:strip_test} returns \true.
\end{proof}

\begin{lemma}
    Let $C$ in $\RR^n$ be a regular curve and $\pi: \RR^n \to \RR^2$ a generic projection. Let $\cA(C)$ be a tubular neighborhood of $C$ constructed with Algorithm \ref{algo:certified_curve_tracking}.  If for two interval regions $I_1$ and $I_2$, $\pi(I_1)$ and $\pi(I_2)$ intersect but their curves do not correspond to a self-intersection, then this is detected after sufficient refinement.
\end{lemma}
    


\begin{proof}
    If there is a nonzero minimum distance between the images of the curves from $I_0$ and $I_1$, then once refinements of $I_0$ and $I_1$ have sufficiently small diameter, their projections will also have small diameter, and they cannot intersect.

    On the other hand, if the minimum distance is zero, this means that $I_1$ and $I_2$ correspond to a single arc of the curve.  After sufficiently many refinements, this portion of the arc is so small that the tangent vector does not change much over the arc and the half-space condition detects no intersections.  
\end{proof}

Putting these lemmas together, we have the following theorem.

\begin{thm}
    Let $\pi:\mathbb{R}^n\rightarrow\mathbb{R}^2$ be a generic projection. For a regular curve $C$ in $\mathbb{R}^n$, Algorithm \ref{algo:certified_plane_curve} returns a tubular neighborhood $\mathcal{A}(C)$ of $C$ such that $\pi(\mathcal{A}(C))$ is a tubular neighborhood of $\pi(C)$. 
\end{thm}
\begin{proof}
We must only argue that in Algorithm \ref{algo:certified_plane_curve}, no intersection is detected twice.  To see this, suppose that an intersection is detected twice.  Since the rectangles $R_i$ for each test of Algorithm~\ref{algo:strip_test} separate the approximations and separate the arcs of the projected curves, the two pairs of boxes must be contained in each other's $R_i$'s, which is not possible by Line 13 of Algorithm~\ref{algo:certified_plane_curve}.  
\end{proof}

\begin{thm}
    \Cref{algo:certified_plane_curve} terminates.
\end{thm}

\begin{algorithm}[ht]
	\caption{CertifiedPlaneCurve}
 \label{algo:certified_plane_curve}
\begin{algorithmic}[1]
\Require  
\begin{itemize}
    \item A regular curve $C=\{c_1,\dots,c_{n-1}\}\subset\mathbb{Q}[X_1,\dots,X_n]$,
    \item a point $x$ approximating a nonsingular point on the curve $C$,
    \item a positive number $r>0$ for the initial radius.
    \item a generic projection $\pi:\mathbb{R}^n\rightarrow \mathbb{R}^2$,
    \item a compact region $D$ in $\mathbb{R}^n$,
    \item constants $\rho\in(0,\frac{1}{2}]$ and $\tau\in(\frac{1}{2},1)$.
\end{itemize}
\Ensure 
\begin{itemize}
    \item a certified tubular neighborhood of a plane curve $\pi(C)$.
\end{itemize}

\State{Compute a tubular neighborhood $\mathcal{A}(C)$ of $C$ using Algorithm $\ref{algo:certified_curve_tracking}$ with $C, x, r, D$, $\tau$, and $\rho$.}
\State{Initialize queue $Q$ containing all pairs of intersecting interval regions $\{ I_1,I_2\}$}
\While{$Q$ is not empty}
\State{Pop pair $\{ I_1, I_2\}$ from $Q$} 
\If{$I_1$ and $I_2$ satisfy the half-space condition}
\State{Accept $I_1$ and $I_2$}
\State{Do not reconsider $I_1$ and $I_2$ or their children as a pair.}
\ElsIf{Algorithm \ref{algo:strip_test} on $I_1$ and $I_2$ returns $\false$}
\State{Refine the interval regions as described in Algorithm \ref{algo:strip_test}.}
\State{Update $Q$ with all pairs of intersecting interval tubes}
\Else{}
\State{Accept $I_1$ and $I_2$}
\State{Discard all other pairs $J_1\subset R_1$ and $J_2\subset R_2$ in $Q$ where $R_1$ and $R_2$ are the regions computed in Algorithm \ref{algo:strip_test}.}
\EndIf
\EndWhile
\State{Return accepted regions.}
 \end{algorithmic}
 \end{algorithm}

To create an approximation to the projection to the curve, we may take the projection of the curve constructed in $\mathbb{R}^n$, with care to remove all self-intersections except those indicated by Algorithm \ref{algo:strip_test}.

 \begin{remark}
If the curve $C$ is not closed and is given by a parametric curve $C=\gamma(t)$.  Then, we may turn this into an implicit system by increasing the dimension by $1$ and using the system of polynomials $X_i-\gamma_i(T)$.  We are interested in projections that include the $T$-direction in their kernel.  If such a projection is generic, then Algorithm \ref{algo:certified_plane_curve} correctly approximates the projection, see Figure~\ref{fig:PTOPO_example}.
 \end{remark}

\section{Implementation details and experiments}\label{sec:implementation}

We provide a proof-of-concept Julia implementation of our algorithms, focusing on the curve tracking algorithm. For interval arithmetic, we use the wrapper of the Arb~\cite{johansson2017arb} library from the package Nemo.jl~\cite{fieker2017nemo}. In curve tracking, we first apply a linear predictor to generate initial position and tangent information, which are required for the Hermite predictor, a variation of the third-order predictor method. We used $\rho =\frac{1}{8}, \tau=\frac{7}{8}, h=\frac{1}{2}$ and $r=\frac{1}{10}$ as the default settings. A GitHub repository holds the code to construct examples (\Cref{fig:tubular_nbd,fig:two_projections,fig:PTOPO_example,fig:MGGJ_example}) in this paper at

\centerline{\url{https://github.com/klee669/certified_curve_projections}}

Figure~\ref{fig:tubular_nbd} illustrates an approximation of a cubic curve along with its tubular neighborhood. It took approximately $600$ iterations to compute. This curve exhibits two points that are close to each other. In contrast, Figure~\ref{fig:two_projections} shows a regular curve, without such near-intersections, where nearby points arise only after projection. The diagrams illustrate that the self-intersection in the tubular neighborhood disappears after sufficient refinement. The curve with $\rho=\frac{1}{8}$ took approximately $300$ iterations, while the curve with $\rho=\frac{1}{32}$ took approximately $550$ iterations. 
\Cref{fig:PTOPO_example}, see \cite[Figure 2]{katsamaki2023ptopo}, required approximately $5600$ iterations to generate the figure. \Cref{fig:MGGJ_example} shows the closed curve from \cite[Section 4.1]{martin2013certified} with nearly singular behavior. It required approximately $2500$ iterations.

\bibliographystyle{abbrv}
\bibliography{ref}

\begin{figure}[h]
    \centering
    \input{graph9}
    \caption{A tubular neighborhood of the curve $C=\{x^3-2.7x-y^2+2\}$ computed with $\rho = \frac{1}{8}$.  Even though the curve track is executed with tilted or curved intervals, rotating interval regions back to the original coordinates with interval arithmetic results in rectangular boxes.}
    \label{fig:tubular_nbd}
\end{figure}

\begin{figure}[h]
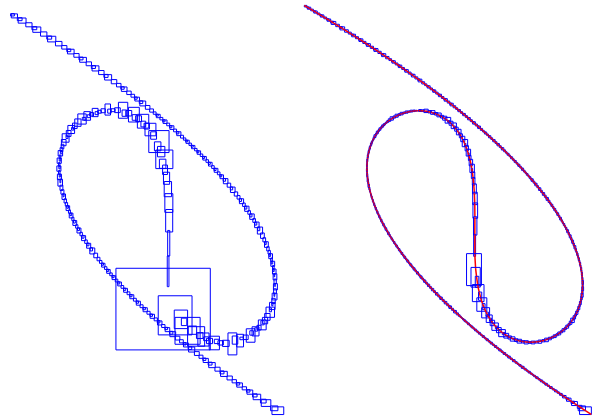

    \centering
     \input{graph_refine_8}
     \input{graph_refine_32}
    \caption{The projection of tubular neighborhoods of $C=\{x+z^5-1.3z^3, y-z^3+z\}$ onto the $xy$-plane. The image on the left is approximated with $\rho=\frac{1}{8}$ and the one on the right is approximated with $\rho=\frac{1}{32}$. The intersecting parts appear with $\rho=\frac{1}{8}$, so the approximation has not been made. The intersecting parts were removed with $\rho=\frac{1}{32}$.}
    \label{fig:two_projections}
\end{figure}



\begin{figure}[h]
    \centering
    \input{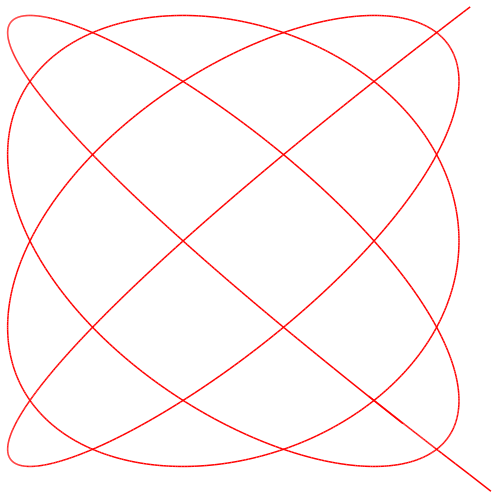}
    \caption{The projection of an approximation of $C=\{x-z^8+8z^6-20z^4+16z^2-2, y-z^7+7z^5-14z^3+7z\}$ from \cite[Figure 2]{katsamaki2023ptopo}}
    \label{fig:PTOPO_example}
\end{figure}

\begin{figure}[h]
    \centering
    \input{MGGJ}
    \caption{The approximation of a closed curve $C=\{x^8-(1-e)x^6+4x^6y^2-(3+15e)x^4y^2+6x^4y^4-(3-15e)x^2y^4+4x^2y^6-(1+e)y^6+y^8\}$ where $e=0.99$ from \cite[Section 4.1]{martin2013certified}}
    \label{fig:MGGJ_example}
\end{figure}

\Addresses

\end{document}